\documentclass[10pt,twocolumn,letterpaper]{article}

\usepackage{wacv}
\usepackage{booktabs}
\usepackage{times}
\usepackage{epsfig}
\usepackage{graphicx}
\usepackage{amsmath}
\usepackage{amssymb}
\usepackage{diagbox}
\usepackage{xcolor}
\usepackage{multirow}
\usepackage{algorithm}
\usepackage{algpseudocode}
\usepackage{tabularx}
\usepackage{soul}
\usepackage{adjustbox}
\usepackage{caption} 
\usepackage{lipsum}
\usepackage{float}
\usepackage{subcaption}
\usepackage{footnote}
\usepackage{enumitem}
\usepackage[sort,square,numbers]{natbib}
\usepackage[font=small,skip=0pt]{caption}

\def\wacvPaperID{0164} 

\wacvfinalcopy 

\ifwacvfinal
\def\assignedStartPage{1} 
\fi

\ifwacvfinal
\usepackage[breaklinks=true,bookmarks=false]{hyperref}
\else
\usepackage[pagebackref=true,breaklinks=true,colorlinks,bookmarks=false]{hyperref}
\fi

\ifwacvfinal
\else
\fi

\begin{document}

\title{OverNet: Lightweight Multi-Scale Super-Resolution with Overscaling Network}

\author{Parichehr Behjati\\
Computer vision center\\
Barcelona, Catalonia Spain \\
{\tt\small pbehjati@cvc.uab.es} 
\and
Pau Rodr\'{i}guez\\
Element AI\\
Montreal, Canada\\  
\and
Armin Mehri\\
Computer Vision Center\\
Barcelona, Catalonia Spain \\
\and
Isabelle Hupont\\
Herta Security\\
Barcelona, Catalonia Spain \\
\and
Carles Fernández Tena\\
Herta Security\\
Barcelona, Catalonia Spain \\
\and
Jordi Gonz\`alez\\
Computer Vision Center\\
Barcelona, Catalonia Spain \\
}

\maketitle


\begin{abstract}
Super-resolution (SR) has achieved great success due to the development of deep convolutional neural networks (CNNs). However, as the depth and width of the networks increase, CNN-based SR methods have been faced with the challenge of computational complexity in practice. Moreover, most SR methods train a dedicated model for each target resolution, losing generality and increasing memory requirements. To address these limitations we introduce OverNet, a deep but lightweight convolutional network to solve SISR at arbitrary scale factors with a single model. We make the following contributions: first, we introduce a lightweight feature extractor that enforces efficient reuse of information through a novel recursive structure of skip and dense connections. Second, to maximize the performance of the feature extractor, we propose a model agnostic reconstruction module that generates accurate high-resolution images from overscaled feature maps obtained from any SR architecture. Third, we introduce a multi-scale loss function to achieve generalization across scales. Experiments show that our proposal outperforms previous state-of-the-art approaches in standard benchmarks, while maintaining relatively low computation and memory requirements.

\end{abstract}

\section{Introduction}

Single image super-resolution (SISR) is the task of reconstructing a high resolution image (HR) from its low-resolution version (LR). As obtaining a HR image from LR is an ill-posed problem, the model needs to \textit{learn} the original data distribution to produce the most likely solutions.  

Convolutional neural networks (CNNs) have recently become the main workhorse to tackle SISR~\cite{dong2014learning}. 
Thanks to the increase in capacity of CNNs in depth~\cite{lim2017enhanced} and width~\cite{yu2018wide}, their performance has greatly improved. Despite their remarkable performance, most deep networks still have some drawbacks. Firstly, increase in depth and width has also raised 
computational demands and memory consumption. This makes modern architectures less applicable in practice, such as in mobile and embedded 
applications. Secondly, as the network depth increases, 
low-level feature information gradually disappears in the successive non-linear operations to produce the output. However, these low-level features are crucial for the network to reconstruct high quality images. 

Aside from the aforementioned problems, another desired ability is to upsample images to arbitrary scales using a single model. Current state-of-the-art SISR models such as RDN~\cite{zhang2018residual}, ESPCNN~\cite{shi2016real} and EDSR~\cite{lim2017enhanced}, only consider SR at certain integer scale factors ($\times2, \times3, \times4$) and treat each super-resolution scale as an independent task. They then train a different specialized model for each, which is not practical for mobile applications. 

To address these problems, we propose Overscaling Network (OverNet), a novel lightweight method for SISR. OverNet consists of two main parts: a lightweight feature extractor and an Overscaling module (OSM) for reconstruction. The feature extractor follows a novel recursive framework of skip and dense connections to reduce low-level feature degradation. The OSM is a new inductive bias which generates accurate SR image by internally constructing an overscaled intermediate representation of the output features. Finally, to solve the problem of reconstruction at arbitrary scale factors, we introduce a novel multi-scale loss by downsampling the output at multiple super resolution factors and we minimize the reconstruction error in all of them. Our main contributions can be summarized as follows: 

\begin{itemize}[leftmargin=*]
\item A lightweight recursive feature extractor, which results in improved performance over 
state-of-the-art models, even those having an order of magnitude more parameters.
\item An Overscaling Module (OSM) that generates overscaled maps from which HR images can be accurately recovered at arbitrary scales. This module boosts the reconstruction accuracy efficiently with respect to its number of parameters. Additionally, we demonstrate that integrating this module into existing state-of-the-art models improves on their original performance.
\item A novel multi-scale loss function for SISR, that allows the simultaneous training of all scale factors using a single model. As a result, the model is able to maintain accurate reconstruction results across scales. 
\end{itemize}

\section{Related Work}

Recently, deep learning models have 
dramatically improved the SISR task. \citet{dong2014learning} first presented SRCNN, a CNN to predict super-resolved images.  
SRCNN has a large number of operations compared to its depth, since the network operates by initially upsampling LR images and subsequently refining them. In contrast to the SRCNN, FSCRNN \cite{dong2016accelerating} and ESPCN \cite{shi2016real} only upsample images at the output of the network, which leads to a reduction in the number of operations compared to SRCNN.  

Despite the higher capacity of deep neural networks, the aforementioned methods have settled for shallow models because of the difficulty in training. VDSR~\cite{kim2016accurate} and IRCNN~\cite{zhang2017learning} improved the performance by increasing the network depth, using stacked convolutions with residual connections. \citet{lim2017enhanced} further expanded the network size and improved the residual block by removing batch normalization layers. 
\citet{ahn2018fast} proposed a cascading residual network using 
ResNet blocks~\cite{he2016deep} to learn the relationship between LR input and HR output. Later, \citet{ledig2017photo} introduced the SRResNet and further improved in \cite{wang2018esrgan} and \cite{tong2017image} by introducing dense connections. More recently, \citet{zhang2018residual} and \citet{liu2020residual} also used dense and residual connections in RDN and RFANet to utilize information from all the feature hierarchy. DBPN \cite{haris2018deep} and SRFBN \cite{li2019feedback} 
architectures comprise of a series of up and down sampling layers 
densely connected with each other. These methods achieved significant improvement over conventional SR methods and indicate the effectiveness of residual learning. 

Another issue of deep learning-based SR is how to reduce the parameters and number of operations to make it effective in mobile applications. For instance, DRCN~\cite{kim2016deeply} was the first to apply recursive algorithm to SISR to reduce the number of parameters by reusing them multiple times.~\citet{tai2017image} improved DRCN by combining the recursive and residual network schemes in order to achieve better performance with even fewer parameters. They also introduced a deep memory network to solve the problem of long-term dependencies~\cite{tai2017memnet}. On the other hand, LapSRN~\cite{lai2017fast} employs a pyramidal framework to increase the image size gradually. By doing so, LapSRN effectively performs SR on extremely low-resolution cases. More recently, \citet{muqeet2020ultra} proposed stacked multi-attention blocks to further improve the performance. However, these methods use very deep networks to compensate for the loss of parameters and hence, they require heavy computing resources. Therefore, we focus on developing a lightweight model to maximize the performance of existing networks as well as minimize their computational cost.

One of the most important stages of SISR is reconstruction, which consists of generating HR images based on high-level features extracted from a low-dimensional space. Interpolation is a commonly used method in SR networks, such as SRCNN~\cite{dong2014learning}, VDSR~\cite{kim2016accurate} and DRRN~\cite{tai2017image},~to resize the LR image to the target size as the input of a CNN model for SR reconstruction. However, 
computational operations are greatly increased due to the large input image size. Thus, FSRCNN \cite{dong2016accelerating} and SRDenseNet \cite{ledig2017photo} directly adopted the LR image as input, 
in which a transposed convolution layer was added to implement the final upsampling reconstruction~\cite{tong2017image}. This method greatly reduces unnecessary computational overhead.~Furthermore, EPSCN \cite{shi2016real} proposed a method called pixelshuffle \cite{aitken2017checkerboard} to overcome the problem of the checkerboard effect in transposed convolution. Pixelshuffle has been widely used in recent 
SR models, such as EDSR \cite{lim2017enhanced}, WDSR \cite{yu2018wide} and RCAN \cite{zhang2018image}. However, these methods cannot manage multi-scale training.

Few works tackle SR at different scale factors, and those that do treat the problem as independent tasks, i.e.~a model is trained for each scale. \citet{lim2017enhanced} proposed the first multi-scale SR model, which has different image processing blocks and upsampling modules for each integer scale factor. Later, \citet{li2018multi} proposed a multi-scale residual network. They use multi-path convolution layers with different kernel sizes to extract multi-scale spatial features.~\citet{grm2019face} proposed to upsample the image  progressively by $\times$2 using a series of so-called SR modules and compute the loss of generated SR results by each module. Thus, these methods require vast amounts of computational resources. Recently, Meta-SR \cite{hu2019meta} introduced an upsampling module based on meta-learning to solve SR at arbitrary scale factors with a single model through a weight prediction technique. However, this method must predict a large number of convolution weights for each target pixel, the prediction is inefficient, and the results may be unstable~\cite{wang2019deep}.

\section{Proposed Overscaling Network}
\begin{figure*}[!t]
    \centering
    \includegraphics[width=0.9\linewidth, height=5.7cm]{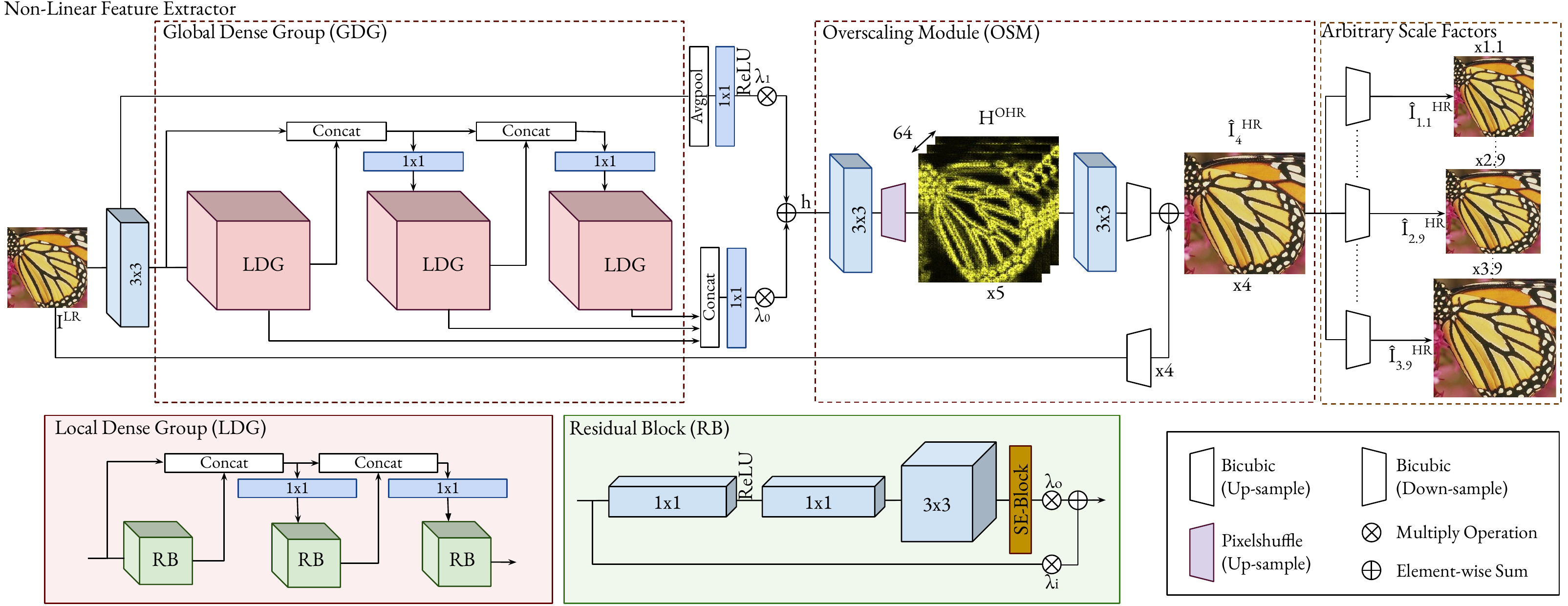}
    \caption{\small Demonstration of our proposed overscaling network with short and long skip connections. As the maximum scale factor in this particular example is set to $N=4$, the required overscaling is $\times$5. }
    \label{fig:architecture}
\end{figure*}

This section describes the main components of our architecture as shown in Figure~\ref{fig:architecture}, and the novel loss function.

\begin{algorithm}[!t]
\caption{\small Overscaling network forward step. Given a LR image and a set of output scales, OverNet produces an HR reconstruction for each scale. Learnable parameters are omitted to improve readability.}
\label{al:overnet}
\begin{algorithmic}
\Function{OverNet}{LR image $I^{LR}$, target scales S}

\State \texttt{\footnotesize \# Compute features with the CNN}
\State $\mathbf{h} = \mathcal{H}(I^{LR})$ 
\State \texttt{\footnotesize \# Overscaling module}
\State $\hat{I}^{HR} = \mathcal{O}(\mathbf{h})$
\State \texttt{\footnotesize \# Output}
\For{s in S}
\State $\hat{I}^{HR}_s = \operatorname{bicubic}_\downarrow(\hat{I}^{HR}, \mathrm{scale}=s)$
\EndFor
\State \Return{\{$\hat{I}_{s}^{HR}, s \in S$\}}
\EndFunction
\end{algorithmic}
\end{algorithm}
\noindent\textbf{Problem formulation.} Algorithm \ref{al:overnet} formulates the main pipeline steps. Given a set of HR images and their downscaled versions $\{I^{HR}, I^{LR}\}$, the goal of SISR is to find a function $\mathcal{F}: LR \rightarrow HR$ that maps LR images to their original HR version. The problem is ill-posed, since there are multiple possible HR images corresponding to a single LR image. However, it is possible to \emph{learn} the most likely reconstruction by parametrizing $\mathcal{F}$ over a set of parameters $\boldsymbol\theta$, and finding the most likely $\boldsymbol\theta$ given some criterion $\mathcal{L}$:

\begin{equation}
    \boldsymbol{\theta}^* = arg \min_{\boldsymbol\theta} \sum \mathcal{L}(\mathcal{F}(I^{LR}, \boldsymbol{\theta}), I^{HR})
\end{equation}

We chose $\mathcal{L}$ to be the $L1$ distance, since we empirically obtained superior PSNR results compared to $L2$. In this work $\mathcal{F}$ is composed of two parts: (i) a feature extractor $\mathcal{H}$: 
\begin{equation}
    \mathbf{h} = \mathcal{H}(I^{LR}, \boldsymbol{\theta}_{h})
\end{equation}
 with parameters $\boldsymbol{\theta}_h$, and (ii) the overscaling module $\mathcal{O}$: 
 \begin{equation}
     \hat{I}^{HR} = \mathcal{O}(\mathbf{h}, \boldsymbol{\theta}_{o})
 \end{equation}
with $\boldsymbol{\theta}_o$ the parameters used in this operation, and $\hat{I}^{HR}$ the reconstructed image. These two parts are described next.

\subsection{Feature Extractor}

The feature extractor computes useful  representations of the LR patch in order to infer its HR version. Concretely, we propose a recursive structure based on {Residual Blocks} (RBs) assembled into {Local Dense Groups} (LDGs) and LDGs into Global Dense Group (GDG), see Figure~\ref{fig:architecture}. 

\noindent\textbf{Residual blocks}.~We use a modified version of WDSR~\cite{yu2018wide} with wide low-rank convolutions instead of using standard residual blocks~\cite{zhang2018residual}. These convolutions widen the activation space before the non-linearity to let more information pass through it and lose less detail, while using the same amount of computation as standard $3\times 3$ residual blocks. In order to make the network focus on more informative features, we exploit the inter-dependencies among feature channels using squeeze-and-excitation ($SE$) operations~\cite{iandola2016squeezenet} after these convolutions, see Figure~\ref{fig:architecture}. 

Inspired by~\cite{szegedy2017inception,srivastava2015highway}, the model learns a scalar multiplier $\lambda$ to balance the amount of information that should be carried by the identity and activation operations within the residual blocks (RBs) of the network.
Let $\textbf{x}_i$ and $\textbf{x}_o$ be the input and output vectors of the $k$-th RB, and $WA$ the wide activation operation~\cite{yu2018wide}. Then, the RB proceeds as:
\begin{equation}
   \textbf{x}_o = \lambda_{o} SE(WA(\textbf{x}_{i})) + \lambda_{i} \textbf{x}_{i}
\end{equation}
\textbf{Local and global dense groups}. RBs are grouped into the so-called Local Dense Groups (LDGs). The input of a RB is concatenated with the output of the all the previous RBs in the group and merged with a 1$\times$1 convolution. This recursion is repeated for all RBs within the LDG. In this way, we gather all local information progressively by 1$\times$1 convolution layers.

To increase the network capacity, a similar recursion is applied to the Global Dense Group (GDG), but this time incorporating skip connections between LDGs. 
We repeat this procedure while integrating the recursive concatenations through the LDGs into a single output. The output of each LDG is concatenated to the input of the next one. In order to facilitate access to local information, the final output of the network receives the concatenation of the outputs of all the LDGs. Therefore, the model incorporates features from multiple layers. This strategy makes information propagation efficient due to the multi-level representation and many shortcut connections. Inspired by MemNet~\cite{tai2017memnet}, we then introduce a 1$\times$1 convolutional layer to adaptively merge the output information, as directly using these concatenated features would greatly increase computational complexity. The output of these hierarchical features can be formulated as
\begin{equation}
    \textbf{f}_{D} = \operatorname{conv}_{1\times1}([\textbf{f}_{0}, ..., \textbf{f}_{D-1}])
\end{equation}
\noindent where $[\textbf{f}_{0}, ..., \textbf{f}_{D-1}]$ refers to the concatenation of feature maps produced by LDGs.

To make sure that no information is lost before the reconstruction step, we incorporate a long-range skip connection to grant access to the original information, and encourage back-propagation of gradients from the output of the feature extractor to the first $3\times 3$ convolution layer. We also include a global average pooling followed by a $1\times1$ convolution, to fully capture channel-wise dependencies from the aggregated information. The final output before the reconstruction step is then, 
\begin{equation}
    \textbf{h} =  \lambda_{0}\textbf{f}_{D} + \lambda_1 \sigma(\operatorname{conv}_{1\times1}(\operatorname{GAP}(\operatorname{conv}_{3\times3}(I^{LR}))))
\end{equation}
where $\sigma$ denotes the ReLU activation, $\operatorname{GAP}$ denotes global average pooling, and $\lambda_0$ and $\lambda_1$ are learned parameters.


\subsection{Overscaling Module}
\label{sec:OSM}
In this work we introduce a new inductive bias in SISR architectures so as to generate images that are more accurate and present fewer artifacts. We hypothesize that, since overscaling produces multiple values for the same pixel, these values act as an ensemble of predictions thus reducing noise when combined to produce the final image. 

Let us consider $N$ the maximum scale factor addressed by the network. 
We first generate an intermediate representation of the final image consisting of overscaled maps $H^{OHR}$, with an overscale factor $(N+1)$ times larger. Thus, given the features $\textbf{h}$ extracted from $I^{LR}$, we use a $3\times3$ convolutional layer followed by the strided sub-pixel convolution proposed in~\cite{aitken2017checkerboard} to upscale the features $\mathbf{h}$ to $H^{OHR}$:

\begin{equation}
    H^{OHR} = \operatorname{pixelshuffle}(\operatorname{conv}_{3\times3}(\mathbf{h}))
\end{equation}

To obtain the final output of the overscaling module, we further include a second long-range skip connection from the original $I^{LR}$ image. The final HR image is obtained by adjusting the overscaled maps and incorporating them into the na\"ive upscaling of the original LR image:
\begin{equation}
    \hat{I}^{HR} = \operatorname{bicubic}_{\downarrow}(\operatorname{conv}_{3\times3}(H^{OHR})) + \operatorname{bicubic}^{\uparrow}(I^{LR})
\end{equation}

Hence, we could think of the whole network as learning how to \textit{refine} or \textit{correct} a na\"ive bicubic upscaling of the low-resolution input, in order to bring it closer to the actual high-resolution counterpart. Since the final $\hat{I}^{HR}$ images are obtained with an efficient non-parametric interpolation, we are able to produce multiple scales with negligible computational cost, and only using differentiable operations. 

\subsection{Multi-Scale Loss}
We propose the minimization of a multi-scale loss to optimize the network. We choose a finite set of scale factors $S = \{s_1 \ldots s_n \}$, all within the interval of scales targeted by the network. 
Once the network has reconstructed the HR image, images at the target scales are obtained through a bank of bicubic interpolators, $\hat{I}_{s}^{HR} = \operatorname{bicubic}_\downarrow(\hat{I}^{HR}, s)$. Then, we minimize the following loss function: 

\begin{equation}
 \mathcal{L} =  \sum_{s \in S}  | \hat{I}^{HR}_{s} - \operatorname{bicubic}_{\downarrow}(I^{HR}, s)|
\end{equation}

Training with this multi-scale loss at different target scales simultaneously provides additional supervision to the model, compared to a single-scale training. As a result, the model is enforced to learn how to generate highly representative overscaled maps, from which HR images at arbitrary scales can be recovered accurately, hence enforcing the generalization capability of the network across scales.

\section{Experimental Results}\label{experiments}


\noindent \textbf{Datasets and metrics}. 
We use the DIV2K dataset for training, a high-quality image dataset containing 800 images for training, 100 for validation and 100 for testing. Several benchmark datasets are used for testing, namely Set5~\cite{bevilacqua2012low}, Set14~\cite{zeyde2010single}, B100~\cite{arbelaez2010contour}, and Urban100~\cite{huang2015single}. SR results are evaluated with two commonly used metrics: PSNR (peak-to-peak signal-to-noise ratio) and SSIM (structural similarity index), on the Y channel of the YCbCr space.\\
\noindent \textbf{Degradation models}. 
To comprehensively illustrate the efficacy of the proposed method, three degradation models are used to simulate LR images, following~\cite{zhang2018learning,zhang2017learning,zhang2018residual}. The first one, denoted by \textbf{BI}, consists of generating LR images by bicubic-downsampling ground truth HR images with $\times$2, $\times$3, $\times$4. The second one, denoted by \textbf{DB}, first performs bicubic downsampling on HR images with $\times$3, and then blurs the images with a Gaussian kernel of size 7$\times$7 and standard deviation 1.6. Finally, we further produce LR images in a third challenging way, denoted by \textbf{DN}, by carrying out bicubic downsampling followed by additive Gaussian noise, with noise level of 30.\\
\noindent \textbf{Implementation details}.  We denote our original model as OverNet and further introduce OverNet w/o OSM (OverNet without overscaling module). We used $64\times64$ RGB input patches from the LR images for training. 
LR patches were sampled randomly and augmented with random horizontal flips and $90^{\circ}$ rotation. The number of LDGs and RBs was set to 3 in all experiments. We trained our models with the ADAM optimizer~\cite{kingma2014adam}. The mini-batch size was set to 64, and the learning rate to the maximum convergent value ${10^{-3}}$, applying weight normalization in all convolutional layers~\cite{yu2018wide}. The learning rate was decreased by half every $2\times{10^5}$ back-propagation iterations. We implemented our networks using the PyTorch framework~\cite{paszke2017automatic} and trained them on NVIDIA 1080 Ti GPUs.

\subsection{Ablation Studies}

To investigate the performance behaviour of the proposed methods we first show how skip connections inside the proposed local and global dense groups affect the performance of OverNet. Next, we analyze the effect of OSM and the multi-scale loss.


 \noindent\textbf{Feature extractor ablation}. Table \ref{tab:DG} presents the ablation study on the effect of skip connections (SCs) inside the local and global dense groups (LDG, GDG). In this work, SCs contain concatenation and 1$\times$1 convolutions. The small changes in number of parameters between columns is due to the removal of SCs with 1$\times$1 convolutions.
\begin{table}[!t]
\centering
\footnotesize
\caption{Effects of skip connections (SCs) in local and global dense groups (LDG, GDG) measured on Urban100 with $\times$3. The best result is \textbf{highlighted}.}
\setlength\arrayrulewidth{0.3pt}

\begin{tabular}{c@{~~~}|c@{~~~~~~~~~}c@{~~~~~~~~~}c@{~~~~~~~~~}c@{~~~~~~}}
\toprule
\multicolumn{1}{c|}{\textbf{Config}} & \textbf{1} & \textbf{2} & \textbf{3} & \textbf{4}\\
\midrule
SCs in LDGs &$\times$&\checkmark&$\times$& \checkmark\\
SCs in GDG &$\times$&$\times$&\checkmark& \checkmark\\\hline
\#Params  & 695K & 806K & 732K & 943K\\
PSNR & 28.23 & 28.21 & 28.29 & \textbf{28.37}\\\bottomrule

\end{tabular}

\label{tab:DG}
\end{table}

\begin{table*}[]
\centering
\footnotesize
\caption{\small PSNR results of different OSM upscaling methods trained for arbitrary scales. The test dataset is B100. Best results are \textbf{highlighted}, second best \underline{underlined}.}
\label{tab:}
\begin{tabular}{c@{~~~~~~~~~~~~}c@{~~~~~~~~~~~~~}c@{~~~~~~~~~~}c@{~~~~~~~~~~}c@{~~~~~~~~~~}c@{~~~~~~~~~~}c@{~~~~~~~~~~}c@{~~~~~~~~~~}c@{~~~~~~~~~~}c@{~~~~~~~~~~}c@{~~~~~~~~~~}}
\toprule
\multirow{2}{*}{\textbf{Experiment}} & \multicolumn{7}{c}{\textbf{Scale}} & \multicolumn{1}{l}{} & \multicolumn{1}{l}{} & \multicolumn{1}{l}{} \\ \cmidrule(l){2-11} 
 & $\times$1.1 & $\times$1.2 & $\times$1.3 & $\times$1.4 & $\times$1.5 & $\times$1.6 & $\times$1.7 & $\times$1.8 & $\times$1.9 & $\times$2.0 \\\hline
Pixelshuffle & 42.40 & 39.71 & 38.10 & 36.75 & 35.60 & 34.70 & 33.96 & 33.30 & 33.65 & 32.22 \\
OSM-bilinear & \underline{42.63} & \underline{39.89} & \underline{38.15} & \underline{36.83} & \underline{35.70} & \underline{34.78} & \underline{34.05} & \underline{33.37} & \underline{32.76} & \underline{32.31}\\
OSM-bicubic &  \textbf{42.74} &  \textbf{39.95} &  \textbf{38.19} &  \textbf{36.87} & \textbf{35.74} & \textbf{34.80} &  \textbf{34.10} &  \textbf{33.42} &  \textbf{32.81} &  \textbf{32.34}  \\\hline
Meta-RDN & \ \underline{42.82} & \ \underline{40.40} & \ \underline{38.28} & \ \underline{36.95} & \ \underline{35.86} & \ \underline{34.90} & \ \underline{34.13} & \ \underline{33.45} & \ \underline{32.86} & \ \underline{32.35}\\
OSM-RDN & \textbf{42.93} & \textbf{40.48} & \textbf{38.42} & \textbf{37.06} & \textbf{36.01} & \textbf{35.02} & \textbf{34.25} & \textbf{35.53} & \textbf{32.95} & \textbf{32.46} \\\hline
 & $\times$2.1 & $\times$2.2 & $\times$2.3 & $\times$2.4 & $\times$2.5 & $\times$2.6 & $\times$2.7 & $\times$2.8 &$\times$2.9 & $\times$3.0 \\\hline
Pixelshuffle & 31.60 & 31.22 & 30.75 & 30.50 & 30.27 & 29.95 & 29.73 & 29.42 & 29.17 & 29.14  \\
OSM-bilinear  & \underline{31.71} & \underline{31.29} & \underline{30.84} & \underline{30.55} & \underline{30.37} & \underline{30.02} & \underline{29.77} & \underline{29.52} & \underline{29.30} & \underline{29.26}    \\
OSM-bicubic & \textbf{31.75}&\textbf{31.34}&\textbf{30.86}&\textbf{30.65}& \textbf{30.42}&\textbf{30.11}&\textbf{29.83}&\textbf{29.64}&\textbf{29.36}&\textbf{29.30}\\
\midrule
Meta-RDN&\textbf{31.82}&\underline{31.41}&\underline{31.06}&\textbf{30.62}&\underline{30.45}&\underline{30.13}&\textbf{29.82}&\underline{29.67}&\textbf{29.40}&\underline{29.30}\\
RDN-OSM & \underline{31.75} &  \textbf{31.46} &  \textbf{31.10} &  \underline{30.60} & \textbf{30.48} & \textbf{30.15} &  \underline{29.79} &  \textbf{29.71} & \underline{29.35} & \textbf{29.38}\\\hline
 & $\times$3.1 & $\times$3.2 & $\times$3.3 & $\times$3.4 & $\times$3.5 & $\times$3.6 & $\times$3.7 & $\times$3.8 & $\times$3.9 & $\times$4.0 \\\hline
 
Pixelshuffle  & 28.78 & 28.70 & 28.50 & 28.30 & 28.14 & 28.10 & 28.72 & 27.74 & 27.60 & 27.65  \\
OSM-bilinear  & \underline{28.81} & \underline{28.77} & \underline{28.62} & \underline{28.49} & \underline{28.23} & \underline{28.22} & \underline{28.90} & \underline{27.82} & \underline{27.79} & \underline{27.75}   \\
OSM-bicubic & \textbf{28.90} & \textbf{28.81} & \textbf{28.66} & \textbf{28.51} & \textbf{28.26}& \textbf{28.25}&\textbf{28.96}&\textbf{27.84}&\textbf{27.83}&\textbf{27.80}\\
\midrule
Meta-RDN&\underline{28.87} & \textbf{28.79} &\underline{28.68} &\underline{28.54}&\underline{28.32} &\textbf{28.27} &\textbf{28.04}&\underline{27.92}& \underline{27.82}&\underline{27.75}\\
RDN-OSM &\textbf{28.96}&\underline{28.70}&\textbf{28.80}&\textbf{28.64}&\textbf{28.41}&\underline{28.23}&\underline{28.00}&\textbf{27.97}&\textbf{27.89}&\textbf{27.83}\\\hline
\end{tabular}
\label{tab:versions}
\end{table*}

It can be observed that the model which used SCs only in GDG attains better performance than the one without SCs (config 1 which is ResNet+OSM) because the short connections inside the GDG effectively carry the information from intermediate to higher layers. Furthermore, by gathering all features before the upscaling module, the model can better leverage multi-level representations.

On the other hand, as discussed in \cite{he2016identity}, multiplicative manipulations such as 1$\times$1 convolutions on the shortcut connection can hamper information propagation, and complicate optimization. Similarly, SCs in LDGs behave as shortcut connections inside the residual blocks. Thus, it is natural to expect performance degradation when the global SCs are deactivated. This is because the global SCs ease the information propagation while the local connections are being learned. Therefore, when OverNet uses SCs in both LDGs and GDG, it outperforms all three models. 

In detail, information propagates globally via SCs used in GDG, and information flows in the LDGs are fused with the ones that come through global connections. By doing so, information is transmitted by multiple shortcuts and thus mitigates the vanishing gradient problem: the advantage of multi-level representation is leveraged by the SCs in GDG, which help the information to propagate to higher layers.
\begin{table*}[htb!]

\caption{\small Average PSNR of SoA methods using OSM instead of their typical upsampling module. The best results are \textbf{highlighted}. }
\begin{adjustbox}{width=1\textwidth}
\footnotesize
\setlength\arrayrulewidth{1.3pt}
\begin{tabular}{c@{~~}c@{~}|c@{~~~}c@{~~~}|c@{~~~}c@{~~~}|c@{~~~}c@{~~~}c@{~~~}|c@{~~~}c@{~~~}}
\toprule
{\textbf{Dataset}} &  {\begin{tabular}[c]{@{}c@{}}\textbf{Scale}\\ \end{tabular}}
 &  
  {\begin{tabular}[c]{@{}c@{}}\textbf{CARN\cite{ahn2018fast}}\end{tabular}} & {\begin{tabular}[c]{@{}c@{}}\textbf{CARN-OSM}\end{tabular}} &
 {\begin{tabular}[c]{@{}c@{}}\textbf{EDSR\cite{lim2017enhanced}}\end{tabular}} & {\begin{tabular}[c]{@{}c@{}}\textbf{EDSR-OSM}\end{tabular}} & {\begin{tabular}[c]{@{}c@{}}\textbf{RDN\cite{zhang2018residual}}\end{tabular}} & 
  {\begin{tabular}[c]{@{}c@{}}\textbf{Meta-RDN\cite{hu2019meta}}\end{tabular}} & {\begin{tabular}[c]{@{}c@{}}\textbf{RDN-OSM}\end{tabular}} &
{\begin{tabular}[c]{@{}c@{}}\textbf{RCAN\cite{zhang2018image}}\end{tabular}} &
 {\begin{tabular}[c]{@{}c@{}}\textbf{RCAN-OSM}\end{tabular}} \\
 

\midrule
{Set5} & {\begin{tabular}[c]{@{}l@{}}$\times$2\\ $\times$3\\ $\times$4\end{tabular}} &  
{\begin{tabular}[c]{@{}c@{}}37.76\\ 34.29\\ 32.13\end{tabular}}&
{\begin{tabular}[c]{@{}c@{}}\textbf{37.90}\\ \textbf{34.35}\\ \textbf{32.15}\end{tabular}}
&{\begin{tabular}[c]{@{}c@{}}38.20\\ 34.76\\ 32.62\end{tabular}} & {\begin{tabular}[c]{@{}c@{}}\textbf{38.28}\\ \textbf{34.80}\\ \textbf{32.66}\end{tabular}} & {\begin{tabular}[c]{@{}c@{}}38.24\\34.71 \\ 32.47\end{tabular}} 
& {\begin{tabular}[c]{@{}c@{}}-\\ -\\ -\end{tabular}} & 
{\begin{tabular}[c]{@{}c@{}}\textbf{38.31}\\ \textbf{34.77}\\ \textbf{32.58} \end{tabular}} &{\begin{tabular}[c]{@{}c@{}}38.27\\ 34.74\\ 32.63 \end{tabular}}  &
{\begin{tabular}[c]{@{}c@{}}\textbf{38.36}\\ \textbf{34.81}\\ \textbf{32.70}\end{tabular}} 
\\ 
\midrule
{Set14} & {\begin{tabular}[c]{@{}l@{}}$\times$2\\ $\times$3\\ $\times$4\end{tabular}} &
{\begin{tabular}[c]{@{}c@{}}33.52\\30.29 \\28.60 \end{tabular}}&
{\begin{tabular}[c]{@{}c@{}}\textbf{33.60}\\\textbf{30.36} \\\textbf{28.68} \end{tabular}}&
{\begin{tabular}[c]{@{}c@{}}34.02\\30.66 \\28.94 \end{tabular}} & {\begin{tabular}[c]{@{}c@{}}\textbf{34.08}\\\textbf{30.71} \\\textbf{29.01} \end{tabular}} & {\begin{tabular}[c]{@{}l@{}}34.01\\30.57 \\28.81 \end{tabular}} & 
{\begin{tabular}[c]{@{}l@{}}34.04\\ 30.55\\28.84 \end{tabular}} & 
{\begin{tabular}[c]{@{}c@{}}\textbf{34.11}\\\textbf{30.63} \\\textbf{28.91} \end{tabular}} &{\begin{tabular}[c]{@{}c@{}}34.12\\30.65 \\28.87 \end{tabular}} &  {\begin{tabular}[c]{@{}c@{}} \textbf{34.19}\\\textbf{30.74} \\\textbf{28.93} \end{tabular}}

\\
 
\midrule
{Urban100} & {\begin{tabular}[c]{@{}l@{}}$\times$2\\ $\times$3\\ $\times$4\end{tabular}} & {\begin{tabular}[c]{@{}c@{}}31.92\\28.06 \\26.07 \end{tabular}} &
{\begin{tabular}[c]{@{}c@{}}\textbf{32.01}\\\textbf{28.12} \\\textbf{26.13} \end{tabular}}&  {\begin{tabular}[c]{@{}c@{}}33.10\\29.02 \\26.86 \end{tabular}} & {\begin{tabular}[c]{@{}c@{}}\textbf{33.15}\\\textbf{29.09} \\\textbf{26.91} \end{tabular}} & {\begin{tabular}[c]{@{}c@{}}32.89\\28.80 \\26.61 \end{tabular}}  &
{\begin{tabular}[c]{@{}c@{}}-\\ -\\- \end{tabular}}  &
{\begin{tabular}[c]{@{}c@{}}\textbf{32.96}\\\textbf{28.91} \\\textbf{26.70} \end{tabular}} &{\begin{tabular}[c]{@{}c@{}}33.34\\ 29.09\\ 26.82\end{tabular}}  &  {\begin{tabular}[c]{@{}c@{}}\textbf{33.40}\\\textbf{29.15} \\\textbf{26.90} \end{tabular}}
 

\\ 
\bottomrule

\end{tabular}
\end{adjustbox}

\label{tab:osm-arch}
\end{table*}

 \noindent\textbf{Effect of the OSM across scales}. Here we analyze the benefits of incoporating the OSM module, and also explore the influence of different interpolation methods on the reconstruction.  {We run the following experiments: (i)~directly using $\operatorname{pixelshuffle}$ to generate the images without overscaling feature maps, followed by bicubic interpolation to downscale to arbitrary scales; (ii) downscaling with bilinear interpolation the overscaled feature maps produced by $\operatorname{pixelshuffle}$ and (iii)~doing the same as (ii) with bicubic interpolation.} As shown in Table \ref{tab:versions}, superior results are achieved by a large margin when the proposed overscaling method is applied. These experiments suggest that, contrary to common practice in the field, the addition of overscaling strongly increases reconstruction accuracy. 
Best results are achieved using OSM with bicubic interpolation, which in turn yields better results than bilinear.

In addition, we compare our results with Meta-RDN~\cite{hu2019meta}, the only method in the literature (to our knowledge) able to carry out SISR at non-integer scales. Meta-RDN is a heavier state-of-the-art model with 22M parameters. For fair comparison, we trained Meta-RDN by replacing its meta-upscale module with OSM (RDN-OSM), while applying their original training settings. RDN-OSM achieves better or comparable performance.

\noindent 
\textbf{OSM across architectures}. The aim of this section is to demonstrate that the benefits of our OSM hold across architectures. 
To this end, we use state-of-the-art networks including CARN~\cite{ahn2018fast}, EDSR\cite{lim2017enhanced}, RDN\cite{zhang2018residual}, Meta-RDN~\cite{hu2019meta} and RCAN\cite{zhang2018image} as references. 
We replaced their typical upsample modules with our overscaling module (CARN-OSM, EDSR-OSM, RDN-OSM and RCAN-OSM in Table~\ref{tab:osm-arch} 
and trained them on DIV2K for all scale factors while applying their original training settings. 

It can be observed that all the methods with OSM have higher PSNR than the corresponding baselines at all scale factors. This shows that OSM is robust and orthogonal to the feature extractor chosen, and it increases PSNR.

\begin{table}[htb!]
\caption{\small Average PSNR to show the performance of OverNet across scales. The test dataset is Set5. Best results are \textbf{highlighted}.}
\centering
\footnotesize
\begin{tabular}{c@{~~~~~}c@{~~~~~}c@{~~~}c@{~~~}c@{~~~}}
\toprule
\multirow{2}{*}[-.3em]{\begin{tabular}{c}\textbf{Overscaling}\\ \textbf{factor}\end{tabular}}& \multirow{2}{*}[-.3em]{\textbf{Parameters}} & \multicolumn{3}{c}{\textbf{Scales}} \\ \cmidrule{3-5} 
& &\multicolumn{1}{l}{$\times$2} & \multicolumn{1}{l}{$\times$3} & \multicolumn{1}{l}{$\times$4}\\
\midrule
$\times$3 & 927K & 38.11 &  --  & -- \\
$\times$4 & 943K & 38.12 & 34.49 & -- \\
$\times$5 & 1079K & 38.14 & 34.54 & 32.32  \\ 
$\times$8 & 955K & \textbf{38.15} & \textbf{34.56} & \textbf{32.36}  \\ \bottomrule
\end{tabular}%

\label{tab:ratio}
\end{table}

\begin{table}[htb!]
\caption{\small Effect of multi-scale loss. OverNet-S uses single-scale loss,  OverNet-M multi-scale loss. Best results are \textbf{highlighted}.}
\centering
\small
\begin{tabular}{l@{~~~~~}c@{~~~~}c@{~~~~}c@{~~~~~~}c@{~~~~}c@{~~~~~}c@{~~~~}}
\toprule
\multirow{2}{*}{\textbf{Dataset}} & \multicolumn{3}{c}{\textbf{OverNet-S}} & \multicolumn{3}{c}{\textbf{OverNet-M}} \\ 
\cmidrule(l){2-4} \cmidrule(l){5-7}
 & \multicolumn{1}{c}{$\times$2} & \multicolumn{1}{c}{$\times$3} & \multicolumn{1}{c}{$\times$4} & \multicolumn{1}{c}{$\times$2} & \multicolumn{1}{c}{$\times$3} & \multicolumn{1}{c}{$\times$4}\\
 \midrule  
Set5 & 38.11 & 34.49 & 32.32 & \textbf{38.23} & \textbf{34.60} & \textbf{32.45} \\ 
B100 & 32.24 & 29.17 & 27.67 & \textbf{32.34} & \textbf{29.30} & \textbf{27.80} \\ 
\multicolumn{1}{l}{Urban100} & 32.44 & 28.37 & 26.31 & \textbf{32.59} & \textbf{28.45} & \textbf{26.42}  \\
\bottomrule
\end{tabular}%

\label{tab:Multi_single}
\end{table}

\noindent \textbf{Generalization across scales}. By construction, the overscaling factor in our architecture is always $(N+1)$ when targeting a maximum scale of $N$, c.f.~Section~\ref{sec:OSM}. The following experiments investigate the generalization capability of models that target a maximum scale $N$ across lower scales \hbox{$M\le N$}. To this end, we trained models for $N\in~\{2,3,4\}$ and evaluated them across scales. Table~\ref{tab:ratio} illustrates the experimental results. It can be observed that models trained to target larger scales yield better PSNR scores for all scale factors. This demonstrates the generalization capabilities of the proposed architecture across scales, as it is not necessary to train a dedicated model for each scale. Instead, training a larger scale seems to be always beneficial for lower scales. Moreover, the cost to pay in terms of additional parameters is low. Note that $\times 4$ and $\times 8$ are composed of multiple consecutive $\times 2$ operations, thus introducing less parameters. Overscaling to higher scales slightly improves the PSNR at the expense of more computation. For the rest of experiments, we overscale to $N+1$ since it still provides significant improvement at slightly higher computational cost.


\noindent \textbf{Effect of multi-scale loss}. Multi-scale learning can process multiple scales with a single trained model, while most of the state-of-the-art algorithms require to train separate models for each supported scale. This property targets real-world applications, where the output size is usually fixed but the input LR scale can vary. Moreover, the multi-scale loss acts as a regularizer, enforcing the generalization of the network across scales and improving performance. As a result, the model is able to maintain accurate reconstruction results across scales. Table \ref{tab:Multi_single} shows experimental results, where the model trained with multi-scale loss achieves better performance with a large margin.

\begin{table*}[!t]

\caption{\small Average PSNR/SSIM values for models with the same order of magnitude of parameters.  Performance is shown for scale factors $\times$2, $\times$3 and $\times$4 with \textbf{BI} degradation. The number of parameters and multi-adds of each method are indicated  under their name. The best performance is shown \textbf{highlighted} and the second best \underline{underlined}. }
\begin{adjustbox}{width=1\textwidth}
\setlength\arrayrulewidth{1.3pt}
\begin{tabular}{c@{~~}c@{~}c@{~~~~}c@{~~~~}c@{~~~~}c@{~~~~}c@{~~~}c@{~~}c@{~~~}c@{~~~}c@{~~~}|c@{~~~}c@{~~~}}
\toprule
{\textbf{Dataset}} &  {\begin{tabular}[c]{@{}c@{}}\textbf{Scale}\\ \end{tabular}}
 & {\begin{tabular}[c]{@{}c@{}}\textbf{VDSR\cite{kim2016accurate}}\\ \textbf{0.7M / 0.6T}\end{tabular}} & {\begin{tabular}[c]{@{}c@{}}\textbf{DRCN\cite{kim2016deeply}}\\ \textbf{1.7M / 18T}\end{tabular}} & {\begin{tabular}[c]{@{}c@{}}\textbf{LapSRN\cite{lai2017deep}}\\ \textbf{0.8M / 30G}\end{tabular}} & {\begin{tabular}[c]{@{}c@{}}\textbf{DRNN\cite{tai2017image}}\\ \textbf{0.3M / 6.8T}\end{tabular}} & {\begin{tabular}[c]{@{}c@{}}\textbf{MemNet\cite{tai2017memnet}}\\\textbf{0.7M / 2.6T}\end{tabular}} &  {\begin{tabular}[c]{@{}c@{}}\textbf{SRFBN\_S\cite{li2019feedback}}\\ \textbf{0.3M / 50G}\end{tabular}} &
{\begin{tabular}[c]{@{}c@{}}\textbf{OISR\_LF\_s\cite{he2019ode}}\\ \textbf{1.4M / 0.26T}\end{tabular}} &
 {\begin{tabular}[c]{@{}c@{}}\textbf{CARN\cite{ahn2018fast}}\\ \textbf{1.6M / 0.2T}\end{tabular}} & 
  {\begin{tabular}[c]{@{}c@{}}\textbf{MAFFSRN-L\cite{muqeet2020ultra}}\\ \textbf{0.8M / 0.1T}\end{tabular}} &
 {\begin{tabular}[c]{@{}c@{}}\textbf{OverNet w/o OSM}\\ \textbf{0.9M / 0.2T}\end{tabular}} & 
 {\begin{tabular}[c]{@{}c@{}}\textbf{OverNet}\\ \textbf{0.9M / 0.2T}\end{tabular}}

  \\

\midrule
{Set5} & {\begin{tabular}[c]{@{}l@{}}$\times$2\\ $\times$3\\ $\times$4\end{tabular}} &  {\begin{tabular}[c]{@{}c@{}}37.53/0.9587\\ 33.66/0.9213\\ 31.35/0.8838\end{tabular}} & {\begin{tabular}[c]{@{}c@{}}37.63/0.9588\\ 33.82/0.9226\\ 31.53/0.8838\end{tabular}} & {\begin{tabular}[c]{@{}c@{}}37.52/0.9591\\ 33.82/0.9227\\ 31.54/0.886\end{tabular}} & {\begin{tabular}[c]{@{}c@{}}37.74/0.9591\\ 34.03/0.9244\\ 31.68/0.8888\end{tabular}} & {\begin{tabular}[c]{@{}c@{}}33.78/0.9597\\ 34.09/0.9245\\ 31.74/0.8893\end{tabular}} & 
{\begin{tabular}[c]{@{}c@{}}37.78/0.9156\\ 34.20/0.9255\\ 31.98/0.8923\end{tabular}} &{\begin{tabular}[c]{@{}c@{}}38.02/0.9605\\ 34.39/0.9272\\ 32.14/0.8947\end{tabular}}  &
{\begin{tabular}[c]{@{}c@{}}37.76/0.9590\\ 34.29/0.9255\\ 32.13/0.8932\end{tabular}} &

{\begin{tabular}[c]{@{}c@{}}38.07/0.9607\\ \underline{34.45/0.9267}\\ 32.20/0.8953\end{tabular}} &

{\begin{tabular}[c]{@{}c@{}}\underline{38.08/0.9609}\\ 34.43/0.9265\\ \underline{32.23/0.8954}\end{tabular}} &

{\begin{tabular}[c]{@{}c@{}}\textbf{38.11}/\textbf{0.9610}\\ \textbf{34.49}/\textbf{0.9267}\\ \textbf{32.32}//\textbf{0.8956}\end{tabular}}
\\
\midrule
{Set14} & {\begin{tabular}[c]{@{}l@{}}$\times$2\\ $\times$3\\ $\times$4\end{tabular}} &  {\begin{tabular}[c]{@{}c@{}}33.05/0.9127\\ 29.78/0.8318\\ 28.02/0.7678\end{tabular}} & {\begin{tabular}[c]{@{}c@{}}33.06/0.9121\\ 29.77/0.8314\\ 28.03/0.7673\end{tabular}} & {\begin{tabular}[c]{@{}c@{}}32.99/0.9124\\ 29.79/0.8320\\ 28.09/0.7994\end{tabular}} & {\begin{tabular}[c]{@{}c@{}}33.23/0.9136\\ 29.96/0.8349\\ 28.21/0.7720\end{tabular}} & {\begin{tabular}[c]{@{}l@{}}33.28/0.9142\\ 30.00/0.8350\\ 28.26/0.7723\end{tabular}} & 
{\begin{tabular}[c]{@{}c@{}}33.35/0.9156\\ 30.10/0.8372\\ 28.45/0.7779\end{tabular}} &{\begin{tabular}[c]{@{}c@{}}33.62/0.9178\\ 30.35/0.8426\\ 28.63/0.7819\end{tabular}} &  {\begin{tabular}[c]{@{}c@{}}33.52/0.9166\\ 30.29/0.8407\\ 28.60/0.7806\end{tabular}} 
&
{\begin{tabular}[c]{@{}c@{}}33.59/0.9177\\ 30.40/0.8432\\ 28.62/0.7822\end{tabular}} &

{\begin{tabular}[c]{@{}c@{}}\underline{33.64/0.9176}\\ \underline{30.41/0.8433}\\ \underline{28.64/0.7823}\end{tabular}}&

{\begin{tabular}[c]{@{}c@{}}\textbf{33.71}/\textbf{0.9179}\\ \textbf{30.47}/\textbf{0.8436}\\ \textbf{28.71}/\textbf{0.7826}\end{tabular}}


\\
 
\midrule
{B100} & {\begin{tabular}[c]{@{}l@{}}$\times$2\\ $\times$3\\ $\times$4\end{tabular}} &  {\begin{tabular}[c]{@{}c@{}}31.90/0.8960\\ 28.83/0.7976\\ 27.29/0.7252\end{tabular}} & {\begin{tabular}[c]{@{}c@{}}31.85/0.8942\\ 28.80/0.7963\\ 27.24/0.7233\end{tabular}} & {\begin{tabular}[c]{@{}c@{}}31.80/0.8949\\ 28.82/0.7973\\ 27.32/0.7264\end{tabular}} & {\begin{tabular}[c]{@{}l@{}}32.05/0.8973\\ 28.95/0.8004\\ 37.38/0.7284\end{tabular}} & 
{\begin{tabular}[c]{@{}c@{}} 32.00/0.8970\\ 28.96/0.8010\\ 27.44/0.7313\end{tabular}} &{\begin{tabular}[c]{@{}l@{}}32.20/0.9000\\ 29.11/0.8085\\ 27.60/0.7369\end{tabular}} & {\begin{tabular}[c]{@{}c@{}}32.08/0.8984\\ 28.97/0.8025\\ 27.44/0.7325\end{tabular}} &  {\begin{tabular}[c]{@{}c@{}}32.09/0.8978\\ 29.06/0.8034\\ 27.58/0.7349\end{tabular}} &

{\begin{tabular}[c]{@{}c@{}}\underline{32.23/0.9005}\\ \underline{29.13/0.8061}\\ 27.59/0.7370\end{tabular}} &

{\begin{tabular}[c]{@{}c@{}}32.18/0.8988 \\ 29.09/0.8033\\ \underline{27.60/0.7371}\end{tabular}}  &

{\begin{tabular}[c]{@{}c@{}}\textbf{32.24}/\textbf{0.9007} \\ \textbf{29.17/0.8063}\\ \textbf{27.67/0.7373}\end{tabular}}  

\\
\midrule
{Urban100} & {\begin{tabular}[c]{@{}l@{}}$\times$2\\ $\times$3\\ $\times$4\end{tabular}} &  {\begin{tabular}[c]{@{}c@{}}30.77/0.9141\\ 27.14/0.8279\\ 25.18/0.7525\end{tabular}} & {\begin{tabular}[c]{@{}c@{}}30.76/0.9133\\ 27.15/0.8277\\ 25.14/0.7511\end{tabular}} & {\begin{tabular}[c]{@{}c@{}}30.41/0.9101\\ 27.07/0.8271\\ 25.21/0.7553\end{tabular}} & {\begin{tabular}[c]{@{}c@{}}31.23/0.9188\\ 27.53/0.8377\\ 25.44/0.7638\end{tabular}} & {\begin{tabular}[c]{@{}c@{}}31.31/0.9195\\ 27.56/0.8376\\ 25.50/0.7630\end{tabular}}  &
{\begin{tabular}[c]{@{}c@{}}31.41/0.9207\\ 26.41/0.8064\\ 24.60/0.7258\end{tabular}} &{\begin{tabular}[c]{@{}c@{}}32.21/0.9290\\ 28.24/0.8544\\ 26.17/0.7888\end{tabular}}  &  {\begin{tabular}[c]{@{}c@{}}31.92/0.9256\\ 28.06/0.8493\\ 26.07/0.7837\end{tabular}}
&
{\begin{tabular}[c]{@{}c@{}}\underline{32.38/0.9308}\\ 28.26/0.8552\\ 26.16/0.7887\end{tabular}} &

{\begin{tabular}[c]{@{}c@{}}32.35/0.9305\\ \underline{28.27/0.8553}\\ \underline{26.22/0.7920}\end{tabular}}&

{\begin{tabular}[c]{@{}c@{}}\textbf{32.44}/\textbf{0.9311}\\ \textbf{28.37}/\textbf{0.8572}\\ \textbf{26.31}/\textbf{0.7923}\end{tabular}}


\\ 
\bottomrule

\end{tabular}
\end{adjustbox}

\label{tab:results}
\end{table*}
\begin{table*}[h!]
\caption{\small Average PSNR/SSIM for models with the same order of magnitude of parameters (RDN included as a high-capacity reference model). 
Scores shown for scale factor $\times3$ using \textbf{BD} and \textbf{DN} degradation models. Best performance is \textbf{highlighted}, second best \underline{underlined}.}
\begin{adjustbox}{width=1\textwidth}

\setlength\arrayrulewidth{1.1pt}

\begin{tabular}{c@{~~}c@{~~}c@{~~~}c@{~~~}c@{~~~}c@{~~~}c@{~~}c@{~~}c@{~~}c@{~~}|c@{~~~}c@{~~~}||c}
\toprule
\textbf{DB} & \begin{tabular}[c]{@{}c@{}} \textbf{Degrad.}\\ \end{tabular} & \begin{tabular}[c]{@{}c@{}} 
\textbf{Bicubic}\\  \end{tabular} & \begin{tabular}[c]{@{}c@{}} \textbf{SPMSR\cite{peleg2014statistical}}\\  \end{tabular} & \begin{tabular}[c]{@{}c@{}} \textbf{SRCNN\cite{dong2014learning}}\\  \end{tabular} & \begin{tabular}[c]{@{}c@{}} \textbf{FSRCNN\cite{dong2016accelerating}}\\  \end{tabular} & \begin{tabular}[c]{@{}c@{}} \textbf{VDSR\cite{kim2016accurate}}\\  \end{tabular} & \begin{tabular}[c]{@{}c@{}} \textbf{IRCNN\_G\cite{zhang2017learning}}\\  \end{tabular} & \begin{tabular}[c]{@{}c@{}} \textbf{IRCNN\_C\cite{zhang2017learning}}\\  \end{tabular} & \begin{tabular}[c]{@{}c@{}} \textbf{SRMD(NF)\cite{tong2017image}}\\  \end{tabular} &
\begin{tabular}[c]{@{}c@{}} \textbf{OverNet w/o OSM}\\ \end{tabular}&
\begin{tabular}[c]{@{}c@{}} \textbf{OverNet}\\ \end{tabular}&
\begin{tabular}[c]{@{}c@{}} \textbf{RDN\cite{zhang2018residual}}\\  \end{tabular} \\ \midrule 
Set5 & \begin{tabular}[c]{@{}c@{}} \textbf{BD}\\ \textbf{DN}\end{tabular} & \begin{tabular}[c]{@{}c@{}}28.34/0.8161\\ 24.14/0.5445\end{tabular} & \begin{tabular}[c]{@{}c@{}}32.21/0.9001\\ -$/$- \end{tabular} & \begin{tabular}[c]{@{}c@{}}31.75/0.8988\\ 28.10/0.7783\end{tabular} & \begin{tabular}[c]{@{}c@{}}26.25/0.8130\\ 24.24/0.6992\end{tabular} & \begin{tabular}[c]{@{}c@{}}33.78/0.9198\\ 27.81/0.7901\end{tabular} & \begin{tabular}[c]{@{}c@{}}33.38/0.9182\\ 24.85/0.7205\end{tabular} & \begin{tabular}[c]{@{}c@{}}29.55/0.8246\\ 26.18/0.7430\end{tabular} & \begin{tabular}[c]{@{}c@{}}34.09/0.9242\\ 27.74/0.8026\end{tabular} &
\begin{tabular}[c]{@{}c@{}} 34.50/0.9270 \\ 28.37/0.8166\end{tabular}
&
\begin{tabular}[c]{@{}c@{}} \textbf{34.59}/\textbf{0.9287} \\ \textbf{28.49}/\textbf{0.8200}\end{tabular} & 
\begin{tabular}[c]{@{}c@{}} \underline{34.58/0.9280}\\ \underline{28.47/0.8151}\end{tabular}\\ \midrule
Set14 & \begin{tabular}[c]{@{}c@{}}\textbf{BD}\\ \textbf{DN}\end{tabular} & \begin{tabular}[c]{@{}c@{}}26.12/0.7106\\ 23.14/0.4828\end{tabular} & \begin{tabular}[c]{@{}c@{}}28.97/0.8205\\ -$/$- \end{tabular} & \begin{tabular}[c]{@{}c@{}}28.72/0.8024\\ 25.55/0.6610\end{tabular} & \begin{tabular}[c]{@{}c@{}}25.63/0.7312\\ 23.10/0.5869\end{tabular} & \begin{tabular}[c]{@{}c@{}}29.90/0.8369\\ 25.92/0.6786\end{tabular} & \begin{tabular}[c]{@{}c@{}}29.73/0.8292\\ 23.84/0.6091\end{tabular} & \begin{tabular}[c]{@{}c@{}}27.33/0.7135\\ 24.68/0.6300\end{tabular} & \begin{tabular}[c]{@{}c@{}}30.11/0.8304\\ 26.13/0.6974\end{tabular} &

\begin{tabular}[c]{@{}c@{}} 30.35/0.8307 \\ 26.56/0.7088\end{tabular} &

\begin{tabular}[c]{@{}c@{}} \underline{30.46/0.8310} \\ \textbf{26.62}/\textbf{0.7116}
\end{tabular} & \begin{tabular}[c]{@{}c@{}}\textbf{30.53}/\textbf{0.8447}\\ \underline{26.60/0.7101}\end{tabular} \\ \midrule
B100 & \begin{tabular}[c]{@{}c@{}}\textbf{BD}\\ \textbf{DN}\end{tabular} & \begin{tabular}[c]{@{}c@{}}26.02/0.6733\\ 22.94/0.4461\end{tabular} & \begin{tabular}[c]{@{}c@{}}28.13/0.7740\\ -$/$- \end{tabular} & \begin{tabular}[c]{@{}c@{}}27.97/0.7921\\ 25.31/0.6351\end{tabular} & \begin{tabular}[c]{@{}c@{}}24.88/0.6850\\ 23.70/0.5856\end{tabular} & \begin{tabular}[c]{@{}c@{}}28.70/0.8003\\ 25.60/0.6455\end{tabular} & \begin{tabular}[c]{@{}c@{}}28.65/0.7922\\ 23.89/0.5688\end{tabular} & \begin{tabular}[c]{@{}c@{}}26.46/0.6572\\ 24.52/0.5850\end{tabular} & \begin{tabular}[c]{@{}c@{}}28.98/0.8009\\ 25.64/0.6495\end{tabular} &

\begin{tabular}[c]{@{}c@{}}29.06/0.8043\\ 25.89/0.6566\end{tabular} &

\begin{tabular}[c]{@{}c@{}} \underline{29.13/0.8060} \\ \textbf{25.95}/\textbf{0.6602}\end{tabular} & \begin{tabular}[c]{@{}c@{}}\textbf{29.23}/\textbf{0.8079}\\ \underline{25.93/0.6573}\end{tabular} \\  \midrule
Urban100 & \begin{tabular}[c]{@{}c@{}}\textbf{BD}\\ \textbf{DN}\end{tabular} & \begin{tabular}[c]{@{}c@{}}23.20/0.6661\\ 21.63/0.4701\end{tabular} & \begin{tabular}[c]{@{}c@{}}25.84/0.7856\\ -$/$- \end{tabular} & \begin{tabular}[c]{@{}c@{}}25.50/0.7812\\ 23.40/0.6590\end{tabular} & \begin{tabular}[c]{@{}c@{}}22.14/0.6815\\ 21.15/0.5682\end{tabular} & \begin{tabular}[c]{@{}c@{}}26.80/0.8191\\ 24.01/0.6802\end{tabular} & \begin{tabular}[c]{@{}c@{}}26.81/0.8189\\ 21.96/0.6018\end{tabular} & \begin{tabular}[c]{@{}c@{}}24.89/0.7172\\ 22.63/0.6205\end{tabular} & \begin{tabular}[c]{@{}c@{}}27.50/0.8370\\ 24.28/0.7092\end{tabular} &

\begin{tabular}[c]{@{}c@{}}28.16/0.8471\\ 24.84/0.7321\end{tabular} &

\begin{tabular}[c]{@{}c@{}} \underline{28.24/0.8485} \\ \textbf{24.93}/\textbf{0.7365} \end{tabular} &  \begin{tabular}[c]{@{}c@{}}\textbf{28.46}/\textbf{0.8582}\\ \underline{24.92/0.7364}\end{tabular}\\  \midrule
\end{tabular}
\end{adjustbox}
\label{tab:BN}
\end{table*}

\subsection{Comparison with State-of-the-art Methods}
\subsubsection{Results with BI degradation models}
We compare the proposed OverNet with nine lightweight state-of-the-art SISR methods~\cite{kim2016accurate,kim2016deeply,lai2017deep,tai2017image,tai2017memnet,li2019feedback,he2019ode,ahn2018fast,muqeet2020ultra}. 
We also train OverNet by replacing its OSM with the typical pixelshuffle upsampling (OverNet w/o OSM). For fair comparison, we train our models individually for each scale factor, including $\times$2, $\times$3 and $\times$4.  We test our models on different benchmarks with PSNR and SSIM.

Table~\ref{tab:results} shows quantitative evaluation results, including the number of parameters and the number of multiplications and additions (multi-adds), for a more informative comparison (under the method name). Multi-adds were calculated with 1280$\times$720 SR images at all scales. Note that, in this table we only compare models that have a roughly similar number of parameters as ours\footnote{Additional analyses and qualitative results can be found as supplementary material.}. OverNet exceeds all the previous methods on numerous benchmark dataset. OverNet w/o OSM also achieves comparable or better results. Results show that both OSM and the proposed feature extractor independently increase PSNR when compared to other SR methods. Finally, combining the proposed feature extractor and OSM together further increases performance.


\setlength\tabcolsep{0.5pt}
\begin{figure}[!t]
\centering
\tiny
\begin{tabular}{ccccc}
\multirow{-6.715}{*}{\adjustbox{}{\includegraphics[width=.35\linewidth, height=2.99cm]{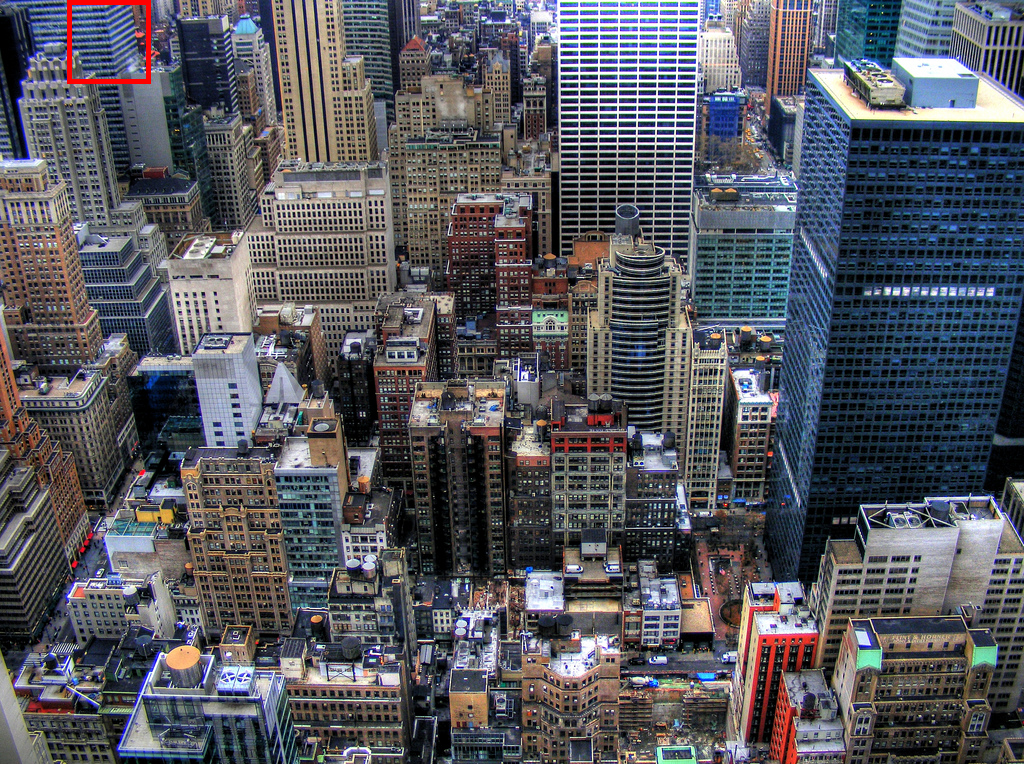}}} 
 & \includegraphics[width=.15\linewidth, height=1.360cm]{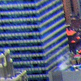} & \includegraphics[width=.15\linewidth, height=1.360cm]{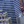} & \includegraphics[width=.15\linewidth, height=1.360cm]{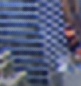} & \includegraphics[width=.15\linewidth, height=1.360cm]{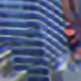} \\
 & HR & Bicubic & VDSR & MemNet \\
 & \includegraphics[width=.15\linewidth, height=1.360cm]{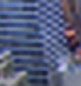} & \includegraphics[width=.15\linewidth, height=1.360cm]{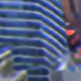} & \includegraphics[width=.15\linewidth, height=1.360cm]{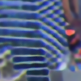} & \includegraphics[width=.15\linewidth, height=1.360cm]{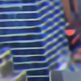} \\
 Img\_073 Urban100 & DRCN & SRFB & CARN &  Ours \\
\end{tabular}

\begin{tabular}{ccccc}
\multirow{-6.715}{*}{\adjustbox{}{\includegraphics[width=.35\linewidth, height=2.99cm]{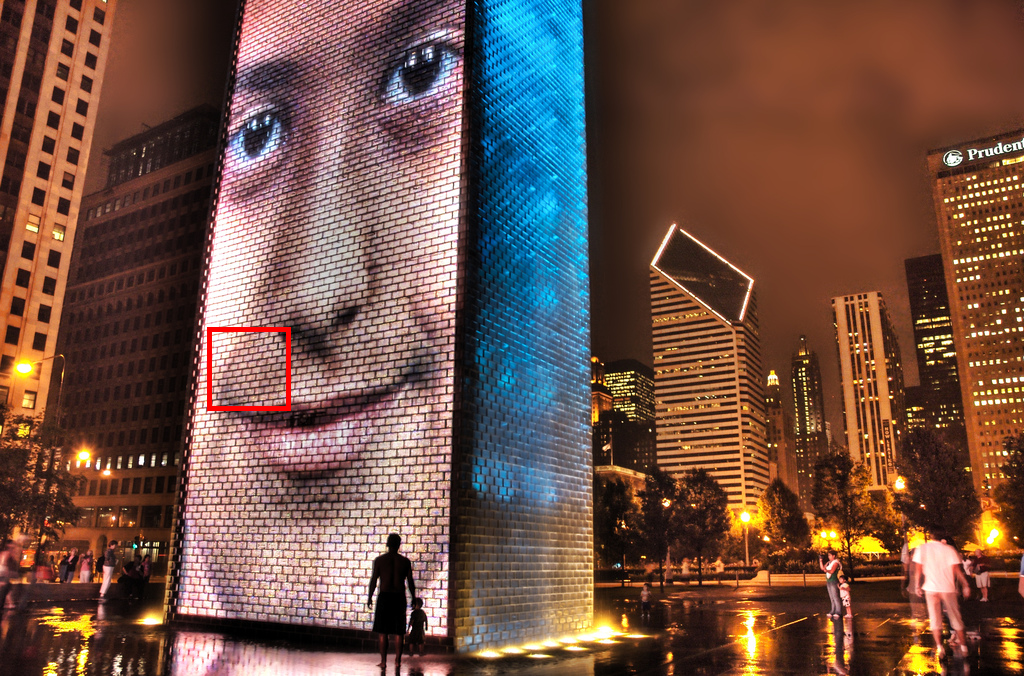}}} 
 & \includegraphics[width=.15\linewidth, height=1.360cm]{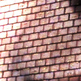} & \includegraphics[width=.15\linewidth, height=1.360cm]{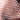} & \includegraphics[width=.15\linewidth, height=1.360cm]{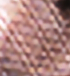} & \includegraphics[width=.15\linewidth, height=1.360cm]{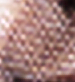} \\
 & HR & Bicubic & VDSR & MemNet \\
 & \includegraphics[width=.15\linewidth, height=1.360cm]{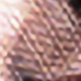} & \includegraphics[width=.15\linewidth, height=1.360cm]{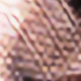} & \includegraphics[width=.15\linewidth, height=1.360cm]{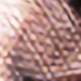} & \includegraphics[width=.15\linewidth, height=1.360cm]{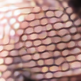} \\
img\_076 Urban100 & DRCN & SRFBN & CARN & Ours \\
\end{tabular}
\caption{\small Visual results of \textbf{BI} degradation model for $\times$4.}
\label{tab:visualize}
\end{figure}

In addition, we present qualitative results in Figure~\ref{tab:visualize}. Our proposal produces high-quality image structures. For image \texttt{Img\_073}, we observe that, unlike OverNet, most of the compared methods fail to recover the definition and orientation of the lines of the blue building. For image \texttt{Img\_076}, the texture of the predicted SR images for all compared methods contains blur or aliasing. In contrast, our proposal partially recovers the brick pattern, resulting in a more faithful SR image.


\setlength\tabcolsep{0.5pt}
\begin{figure}[!t]
\centering

\tiny
\begin{tabular}{ccccc}
\multirow{-6.716}{*}{\adjustbox{}{\includegraphics[width=.3\linewidth, height=2.99cm]{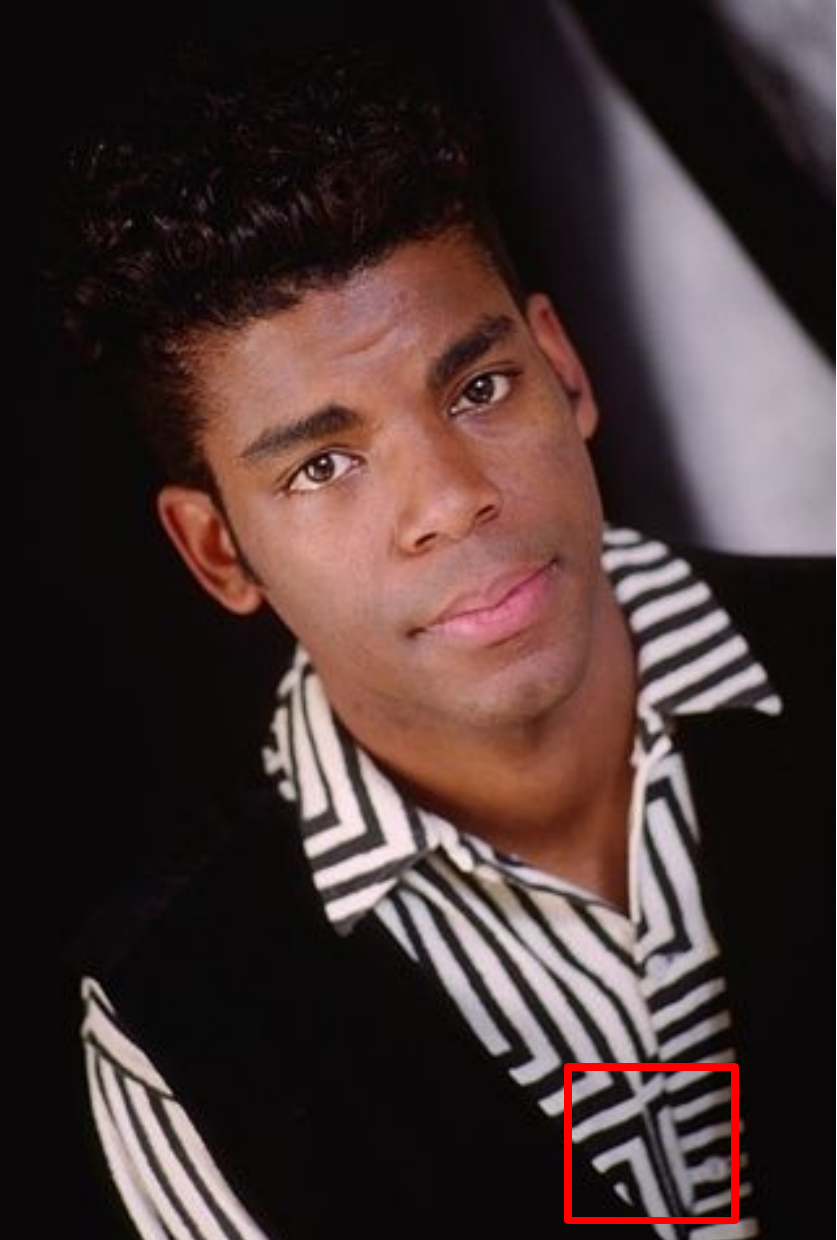}}} 
 & \includegraphics[width=.16\linewidth, height=1.400cm]{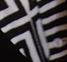} & \includegraphics[width=.16\linewidth, height=1.400cm]{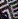} & \includegraphics[width=.16\linewidth, height=1.400cm]{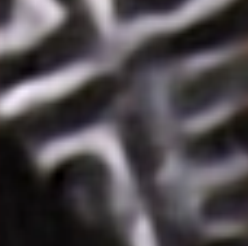} & \includegraphics[width=.16\linewidth, height=1.400cm]{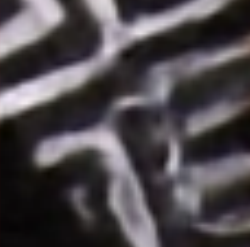} \\
 & HR & Bicubic & SRCNN & VDSR \\
 & \includegraphics[width=.16\linewidth, height=1.400cm]{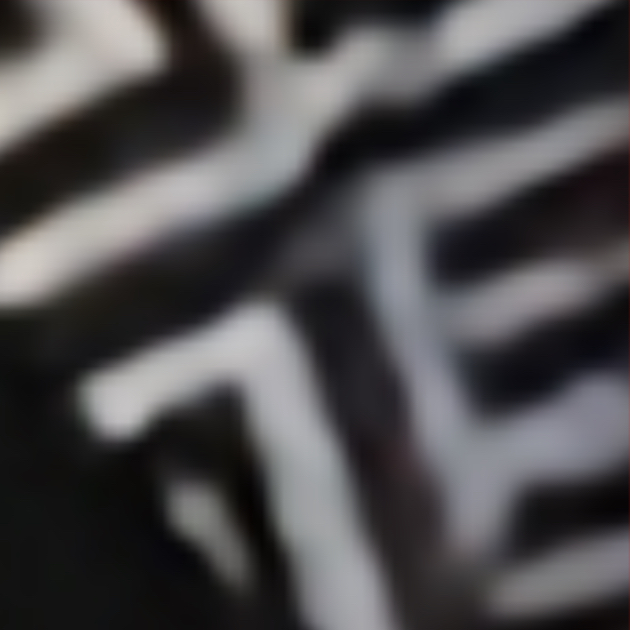} & \includegraphics[width=.16\linewidth, height=1.400cm]{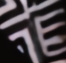} & \includegraphics[width=.16\linewidth,height=1.400cm]{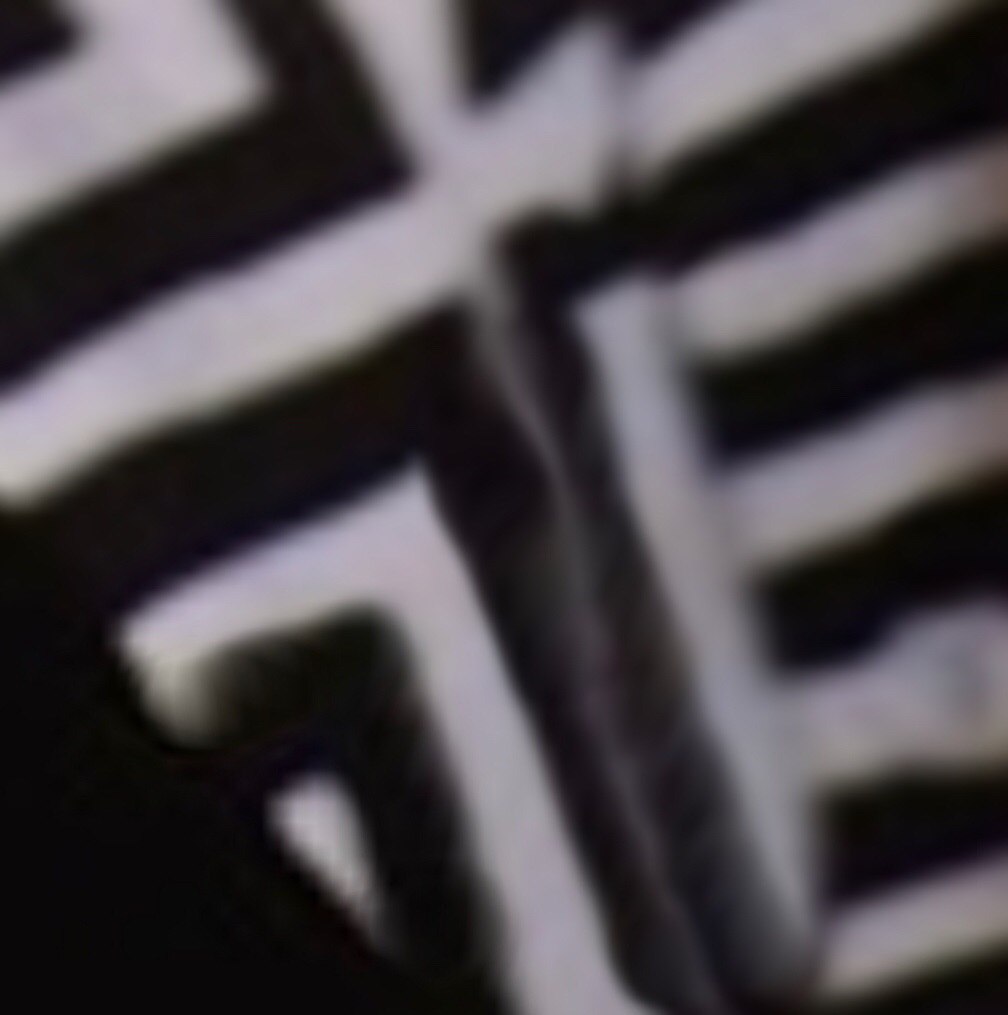} & \includegraphics[width=.16\linewidth, height=1.400cm]{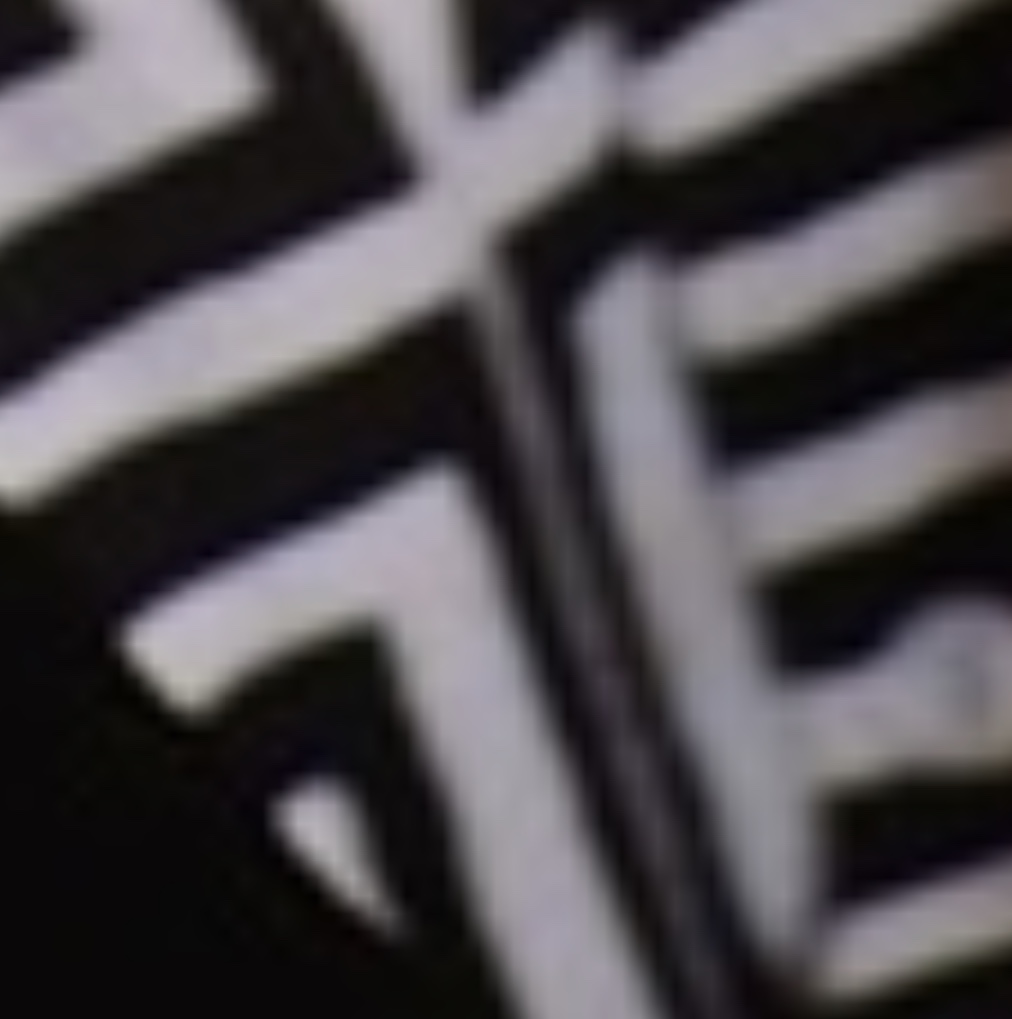} \\
Img\_063 B100 (BD)& IRCNN & SRMD & RDN & Ours \\
\end{tabular}

\begin{tabular}{ccccc}
\multirow{-6.716}{*}{\adjustbox{}{\includegraphics[width=.3\linewidth, height=2.99cm]{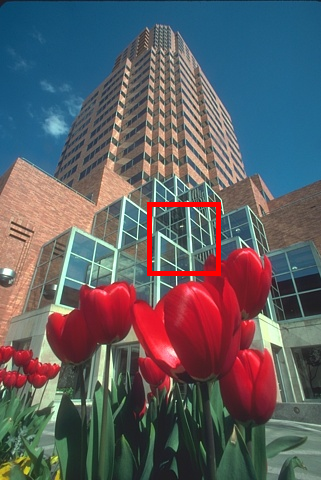}}} 
 & \includegraphics[width=.16\linewidth, height=1.400cm]{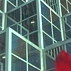} & \includegraphics[width=.16\linewidth, height=1.400cm]{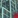} & \includegraphics[width=.16\linewidth, height=1.400cm]{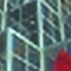} & \includegraphics[width=.16\linewidth, height=1.400cm]{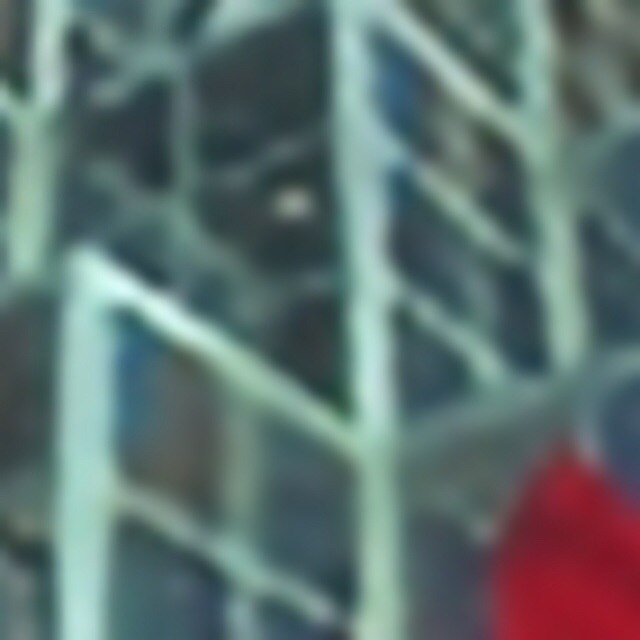} \\
 &HR & Bicubic & SRCNN & FSCRNN \\
 & \includegraphics[width=.16\linewidth, height=1.400cm]{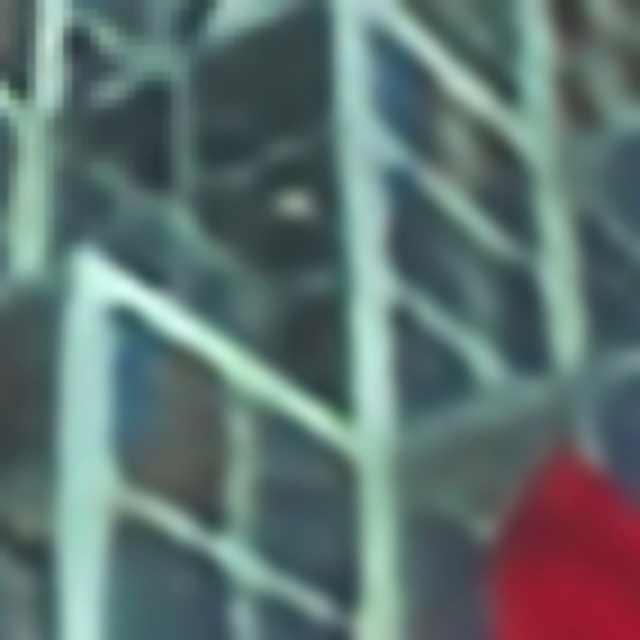} & \includegraphics[width=.16\linewidth, height=1.400cm]{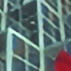} & \includegraphics[width=.16\linewidth, height=1.400cm]{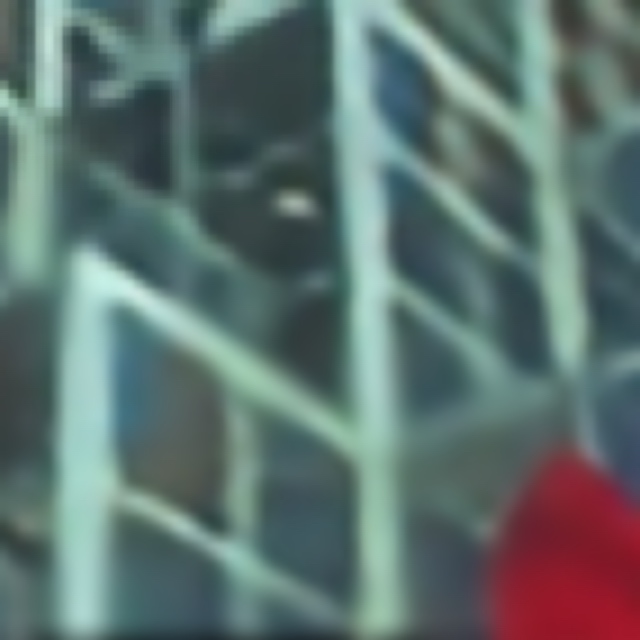} & \includegraphics[width=.16\linewidth, height=1.400cm]{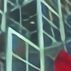} \\
Img\_095 B100  (DN) & VDSR & IRCNN\_G & SRMD & ours  \\
\end{tabular}
\caption{\small Visual results of \textbf{BD} and \textbf{DN} degradation models for $\times$3. }
\label{tab:noisy_visualize}
\end{figure}

\subsubsection{Results with BD and DN degradation models}\label{degradation}

Following~\cite{zhang2018residual}, we show the results obtained after applying \textbf{BD} and \textbf{DN} degradation models, and compare to seven SR methods \cite{peleg2014statistical,dong2014learning,dong2016accelerating,kim2016accurate,zhang2017learning,tong2017image}, see Table~\ref{tab:BN}. We included the RDN~\cite{zhang2018residual} high-capacity model for reference. Because of the mismatch of degradation setups, SRCNN~\cite{dong2014learning}, FSRCNN~\cite{dong2016accelerating}, and VDSR~\cite{kim2016accurate} have been re-trained for both BD and DN. 
Our models achieve the best PSNR and SSIM scores over other SR methods with similar capacity. It can be observed that RDN performs slightly better in some BD datasets but not in DN datasets. Thanks to OSM, OverNet is able to reduce the DN degradation to obtain better results when compared to RDN. It is worth noting that while RDN has 22M parameters, OverNet only has 0.9M parameters.


In Figure~\ref{tab:noisy_visualize} we show two sets of visual results with the \textbf{BD} and \textbf{DN} degradation models from the standard benchmark datasets. For \textbf{BD} degradation, other methods were unable to remove blurring artifacts. In contrast, OverNet could alleviate distortions and generate more accurate details in the SR images. Regarding \textbf{DN} degradation, we observe that it is difficult to recover the details with the other methods. However, our method can deliver good results by removing more noise and enhancing details.

\begin{figure}[!t]
    \centering
    \includegraphics[width=1\linewidth, height=6cm]{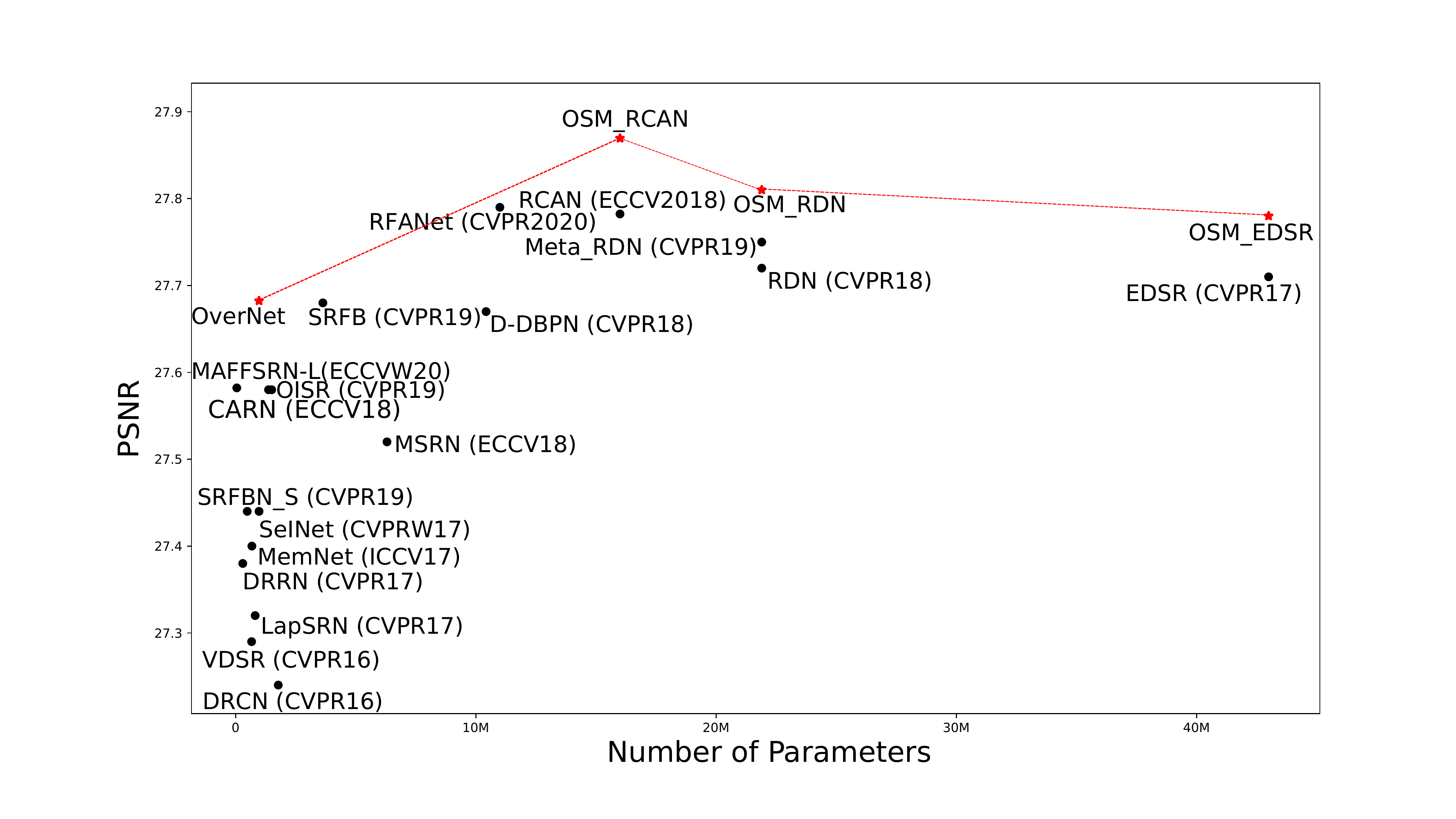}
\caption{\small Comparative capacity and performance of state-of-the-art SISR models. The \textcolor{red}{red stars} represents our methods.}
    \label{fig:param}
\end{figure}

\subsubsection{Memory complexity and running time analysis}
In Figure~\ref{fig:param}, we compare OverNet against various benchmark algorithms in terms of network parameters and reconstruction PSNR, using the B100 dataset with a scale of $\times$4. OverNet achieves the best SR results among all the lightweight SR networks with fewer parameters. In comparison with the networks with a large number of parameters, our proposed OverNet achieves better or competitive results. This demonstrates our method can well balance the number of parameters and the reconstruction performance. We also replace the original upsample modules from different SR methods with OSM: RDN, EDSR and RCAN (RDN+OSM, EDSR+OSM and RCAN+OSM). It can be observed that all the methods with OSM have higher PSNR than the corresponding baselines.


We compare the running time of OverNet on Urban100 with five other state-of-the-art networks, namely MemNet~\cite{tai2017memnet}, EDSR~\cite{lim2017enhanced}, SRFBN~\cite{li2019feedback}, D-DBPN~\cite{haris2018deep}, and Meta-RDN~\cite{hu2019meta}, using a scale factor $\times4$. The running time of each network is evaluated using its official code, on the same machine with a NVIDIA 1080 Ti GPU. OverNet is the fastest (see Table \ref{tab:eff}), reflecting its efficiency.
\begin{table}[!t]
\footnotesize
\caption{\small Average running time comparison on Urban100 for $\times 4$.}
\centering
\begin{tabular}{l@{~~~~~}c@{~~~~~~~}c@{~~~~~~~}c@{~~~~~~~}}
\toprule
\multirow{1}{*}[-.3em]{\begin{tabular}{c}Model\end{tabular}} & \multirow{1}{*}[-.3em]{Parameters}& \multirow{1}{*}[-.3em]{Running Time (s)} & \multirow{1}{*}[-.3em]{PSNR} \\  
\midrule
MemNet & 0.6M & 0.481 & 25.54  \\
EDSR & 43M&1.218 & 26.64 \\
SRFBN\_S & \textbf{0.4M} & 0.006 & 25.71 \\
D-DBPN & 10M & 0.015 & 26.38\\
RDN & 22M & 1.268 & 26.61\\
Meta-RDN & 22M & 1.350 & \textbf{26.65}\\
\textbf{Ours} & 0.9M &\textbf{0.004} & 26.31\\ \bottomrule
\end{tabular}%

\label{tab:eff}
\end{table}




\section{Conclusion}
We introduced OverNet, a novel efficient architecture for image super-resolution at arbitrary scales using a single model. OverNet outperforms state-of-the-art algorithms with a reduced number of parameters and low computational requirements. The main contributions are: (i) a lightweight feature extractor that enhances the flow of information to preserve details; (ii) an Overscaling Module that helps to generate accurate SR images at different scaling factors, and (iii) a multi-scale loss that improves training compared to dedicated single-scale models. Thanks to the OSM, we can train a single model for super-resolution at arbitrary scale factors. We proved that the overscaling head can be flexibly applied to other SR models by simply replacing their upsampling module, thus improving their original performance. The provided evidence suggests that the proposed overscaling method may help with other low-level image restoration tasks, such as denoising and dehazing.

\noindent \textbf{Acknowledgements}. Authors acknowledge the funding received by the Spanish project TIN2015-65464-R (MINECO/FEDER).


{\small
\bibliographystyle{abbrvnat}
\bibliography{egbib}
}

\end{document}